\documentclass[superscriptaddress,twocolumn,amsmath]{revtex4}

\usepackage{graphicx,color,url}
\definecolor{b}{rgb}{0,0,1.0}
\definecolor{r}{rgb}{1,0,0}
\definecolor{g}{rgb}{0,1,0}

\begin{document}

\newcommand{\HASBUTE}{HAS-BUTE Condensed Matter Research Group, 
Budapest University of Technology and Economics, H-1111 Budapest, Hungary}

\newcommand{\RR}{\color{r}}
\newcommand{\GG}{\color{g}}
\newcommand{\BB}{\color{b}}
\newcommand{\Phicrit}{{\Phi_{\text{crit}}}}
\newcommand{\taucrit}{{\tau_{\text{crit}}}}
\newcommand{\mucrit}{{\mu_{\text{crit}}}}
\newcommand{\mucont}{{\mu_{\text{cont}}}}
\newcommand{\mueff}{{\mu_{\text{eff}}}}
\newcommand{\muup}{{\mu_{\text{up}}}}
\newcommand{\mulo}{{\mu_{\text{lo}}}}
\newcommand{\Lshear}{\lambda_{\text{shear}}}
\newcommand{\gammadot}{\dot\gamma}
\newcommand{\vshear}{{v_{\text{shear}}}}
\newcommand{\Alo}{{A_{\text{lo}}}}
\newcommand{\Aup}{{A_{\text{up}}}}

\title{Collective rheology in quasi static shear flow of granular media}

\author{Tam\'as Unger}

\email{unger@phy.bme.hu}
\affiliation{\HASBUTE}

\begin{abstract}
This paper is devoted to the basic question of what factors determine the
strain field in a quasi static granular flow. It is shown that using stress -
strain rate relations is not the proper way to understand quasi static
rheology. An alternative approach is discussed where the local deformation 
is regarded as the cause of deformation in the vicinity. We suggest a
continuum model 
where the local shear strain is proportional to its Laplacian and the
proportionality factor is determined by the local stress. The predicted
behavior is tested in a three dimensional shear cell by means of
computer simulations. The simplicity of our setup makes it ideal to
demonstrate and examine the fundamental open questions of collective
granular flows. The observed shear profile is interpreted in the
framework of the suggested model.
\end{abstract}

\maketitle

\section{Introduction}

When a granular material is deformed slowly in a simple plane shear cell
\cite{GDRMiDi04,Craig04,Radjai04} a critical state
is reached after a transient period. This critical state is unique for the
given granular material and is characterized e.g. by a well defined
critical volume fraction $\Phicrit$ and a critical coefficient of effective
friction $\mucrit$. A nice property of the critical state is that it is
devoid of memory effects in the sense that the preparation history does not
play any role \cite{Craig04,Radjai04}. The initial state of the contact
network is destroyed and 
a self-organized inner fabric is maintained during the shear deformation.

The material constants $\Phicrit$ and $\mucrit$ are valid for quasi static
shear of low inertia numbers $I$ \cite{GDRMiDi04}, i.e. when the shear rate
is low enough and the pressure is large enough. In this paper we
concentrate on the quasi static regime rather than on dense inertia
flows. The dilute gas-like regime is out of scope of the present study.

The idea in the plane shear experiment is that the
material is tested under uniform stress and strain conditions in order to
deduce constitutive relations between local strain, strain rate and
stress. Then with this knowledge one can try to understand the rheology in
more complicated situations that often come up in applications. It is
expected that a small homogeneous piece of bulk material of a large system
behaves according to the plane shear experiment. One of the most
interesting things about quasi static deformations is that the above
concept does not work.

The meaning of the effective friction coefficient is the ratio of the shear
stress measured in the shear direction divided by the perpendicular normal
stress. What we learn from plane shear tests performed both in experiments
and in computer simulations is that persistent shear flow needs a stress
ratio at least $\mucrit$. Once the imposed stress ratio drops below this
threshold the flow stops and the material behaves as a solid body.

On the other hand if a bit more complicated setup is considered, e.g. a
Couette cell \cite{GDRMiDi04}, the above rule ``solid-below-the-threshold''
does not hold any more. The problem is more pronounced for split bottom
shear cells \cite{Fenistein04,Unger04a,Depken06,Cheng06} where wider shear
zones and smoother flow profiles arise than in Couette flows. In these
systems stationary shear flow is observed also in regions where the local
stress ratio is far below the threshold $\mucrit$
\cite{Ries07,Depken07}. An extreme example is shown in \cite{Nichol10}
where sand behaves as a zero yield stress fluid, i.e. even a tiny shear
stress is enough to cause deformation.

We conclude that such a collective flow can not be interpreted locally as a
simple plane shear. Clearly, the local strain rate is not determined by the
local stress tensor alone. What is then the underlying mechanism that
determines the stress and strain fields? How does the material know locally
how to deform? This is the basic mystery of quasi static shear flow that is
in the focus of the present paper. To answer these questions is of
fundamental importance in order to be able to predict rheology in various
granular systems. Even a very simple geometry such as the straight split
bottom cell \cite{Ries07,Depken06} represents a serious challenge for
theories. We are not aware of any model in the granular literature that is
able to provide the correct flow profile for this case. Let us refer to the
above questions as the problem of collective rheology.

A popular approach to describe the flow is to assume the effective friction
coefficient to depend on the local inertia parameter $\mueff(I)$ and indeed
it does a very good job for inertia flows \cite{Jop06,GDRMiDi04}. This
technique fails, however, for quasi static flows. It can not describe the
non-trivial smooth flow that emerges in the collective rheology and often
predicts infinitely narrow shear bands instead \cite{Jop08,Unger04a}.

To study the problem of collective rheology it is a good start to deal with
stationary flows for the sake of simplicity. Thus Couette or split bottom
tests seem to be good choices. However, these systems are much more
complicated than the plane shear experiment, they involve too many degrees
of freedom and unnecessary difficulties. Just to mention one, the material
layers that slide next to each other during the flow are curved in one way
or another, which affects the stress field in a non-trivial way
\cite{Depken06} and makes the treatment harder. To shed more light on the
mystery it is better to start with the simplest possible system where the
problem of collective rheology arises. In this article we suggest and
analyze such a test system, which is only one step away from the
homogeneous plane shear.

The shear flows investigated here are achieved inside the computer by means
of discrete element simulations. We deal with three dimensional flows of
frictional and spherical grains. The paper is structured as follows. First
we examine the important reference situation of the quasi static critical state
produced in homogeneous plane shear. Then we modify the system to
achieve non-trivial collective flow and present the results of the
simulations. In the last part we suggest a model that can account for some of
the features of this collective behavior.

\section{Homogeneous plane shear}

\subsection{The influence of the inertia parameter}

\label{inertiasection}

The computer simulations are carried out by the algorithm of contact
dynamics (CD) \cite{Jean99,Moreau94,Brendel04,Shaebani09}. The dynamics, as
usual, is governed by Newton's equation of motion applied for each
grain. The interaction of the grains are handled by means of constraint
forces at the contact points which are determined based on the assumption
that the grains are rigid (undeformable).

\begin{figure}[tb]
\includegraphics[scale=0.35]{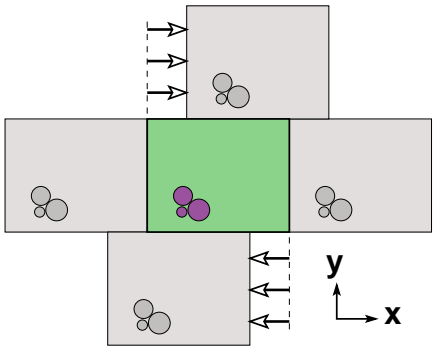}
\caption{Schematic illustration of the shear cell used in the computer
  simulations. The shear cell is three dimensional where Lees-Edwards
  boundary condition is applied in $y$ direction while in $x$ and $z$ directions
  normal periodic boundaries are used.
  \label{LeesEdwards}}
\end{figure}

Our shear cell uses Lees-Edwards boundary condition in order to attain
homogeneous plane shear (see Fig.~\ref{LeesEdwards}).
That is the system is periodically connected in $y$ direction in a special
way: There is a velocity difference $\vshear$ between the upper and lower
boundary which provides the shearing. In $x$ and $z$ directions normal
periodic boundaries are used. Thus confining walls are completely avoided
and shearing is achieved by the special periodic boundaries alone. This is
advantageous because walls would introduce undesired boundary effects
(e.g. layering) that significantly can alter the behavior especially of
small systems used in computer simulations. This way the system is a priory
translational invariant in $x$, $y$ and $z$ directions, there are no
special points, everywhere is ``inside the bulk''. The pressure conditions
are controlled by Andersen boundary conditions \cite{Andersen80}, i.e. by
properly varying the volume. Details how this is implemented in CD can be
found in \cite{Shaebani09}. In all the simulations gravity is set to zero.

Thus our system is sheared in the $xy$ plane in $x$ direction and the two
following parameters are kept constant during the simulation: the global
shear rate $\gammadot=\vshear / (2 L_y)$, where $L_y$ is the system size
in $y$ direction, and the $yy$ component of the stress
tensor ($\sigma_{yy}$). All simulations presented in this paper are
performed with the help of the above shear cell.

In the first set of simulations we examine one system in various
conditions. This test system consists of $2500$ spherical grains with radii
distributed uniformly between the unit length $1.0$ and $1.3$. The friction
coefficient $\mucont$ of the grain-grain contacts is set to 0.2 (for
both static and dynamic friction). We are interested in the steady state
properties therefore relatively large shear displacement is simulated for
each run. The value of the total shear strain $\gamma = \Delta
l_{\text{shear}} / (2 L_y)$ (given by half of the total shear displacement
in $x$ direction divided by the system size in $y$ direction) is typically of
the order of $100$. Even for the smallest inertia number where the
simulation is very computation consuming $\gamma$ is chosen larger
than $20$.

We examine whether the inertia parameter $I$ \cite{GDRMiDi04} serves as a
good control parameter that determines the state of the system. The inertia
parameter is defined as
\begin{equation}
I=\gammadot d \sqrt{\rho/p} \, ,
\end{equation}
where $d$, $\rho$ and $p$ are the typical grain size, the mass density of
grains and the pressure, respectively. The inertia parameter reflects the
scale of the inertia forces with respect to the force scale generated by
the pressure. Low inertia parameter means that the inertia forces that
cause accelerations of the grains are much smaller than the typical contact
forces between the grains. The definition of $p$ is ambiguous because the
stress tensor is not spherical. Here we use $\sigma_{yy}$ in place of $p$,
however, this choice is unimportant in the scope of the present
study. Using any of the normal stress components or $\text{Tr}(\sigma)/3$
would lead to negligible change in the value of $I$.

We focus basically on two indicators of the state of the material, the
density and the resistance against shear. More precisely, these are the
solid fraction $\Phi$ which gives the total solid volume of the grains
divided by the volume of the system and the effective friction $\mueff$
provided by the ratio of two elements of the stress tensor $\sigma_{xy} /
\sigma_{yy}$. We measure the time averaged $\Phi$ and $\mueff$ in the
stationary flow and plot these data as the function of the inertia parameter. 

\begin{figure}[tb]
\includegraphics[scale=0.65]{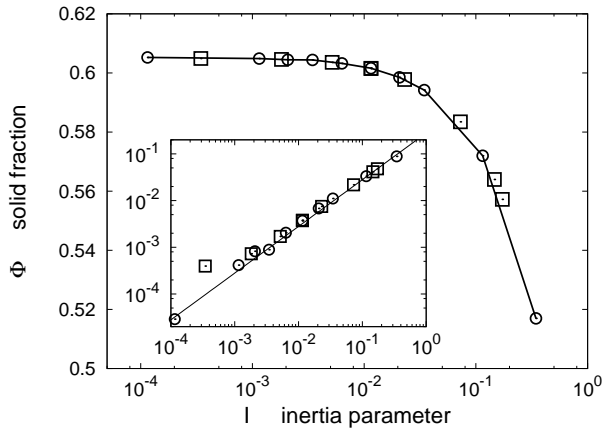}
\caption{The solid fraction $\Phi$ is plotted against the inertia parameter
  $I$. The circles connected by the solid line show a set of
  simulations which differ only in the driving shear rate $\gamma$, while
  squares stand for simulations where other parameters are also varied in a
  random manner such as the mass density of grains, the pressure and the
  technical parameter $\Delta t$ which is the time step of the
  algorithm. The inset shows the difference from the critical value
  ($\Phicrit - \Phi(I)$) against $I$ in log-log scale. The straight line
  indicates slope $1$.
  \label{Phi}}
\end{figure}

In Fig.~\ref{Phi} the results for the solid fraction are shown.  We scanned
through a wide region of $I$ by systematically changing the driving shear
rate $\gamma$, while keeping all other input parameters fixed. Other
simulations were also performed where we tested several random combinations
of the input parameters: The pressure and mass density of the grains
were changed by factors over $1000$, the time step used by the simulation
code was also varied by a factor $50$. It can be seen that putting all
theses data in one plot make them collapse on a single curve. The influence
of other physical and technical parameters (not shown here) were also
tested: The size of the system \footnote{The sizes $L_x$, $L_y$ and $L_z$
  were varied one by one in the range from $10$ to $60$ and the total
  number of grains were changed accordingly. These tests left the solid
  fraction and the stress tensor unchanged.}, the inertia used by the
Andersen type of pressure control and the number of force-iterations
\cite{Brendel04} used by the contact dynamics algorithm are all parameters
that within a wide range have no effect on the value of $\Phi$.

\begin{figure}[tb]
\includegraphics[scale=0.65]{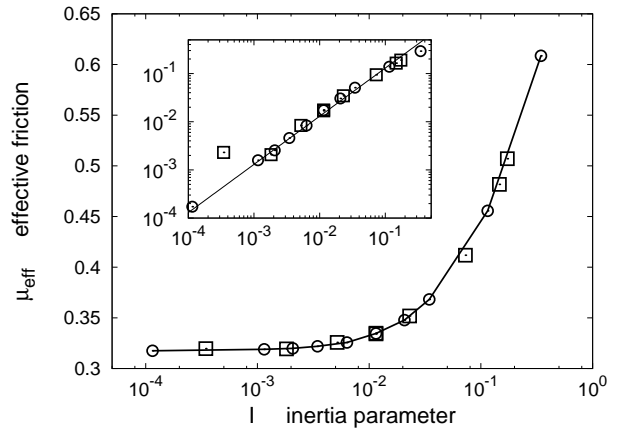}
\caption{The effective friction $\mueff$ is plotted against the inertia
  parameter $I$. The symbols (circles and squares) have the same meaning as
  in Fig.~\ref{Phi}. The inset shows the deviation from the critical value
  ($\mueff(I) - \mucrit$) against $I$ in log-log scale. The straight line
  indicates slope $1$.
  \label{mueff}}
\end{figure}

\begin{figure}[tb]
\includegraphics[scale=0.65]{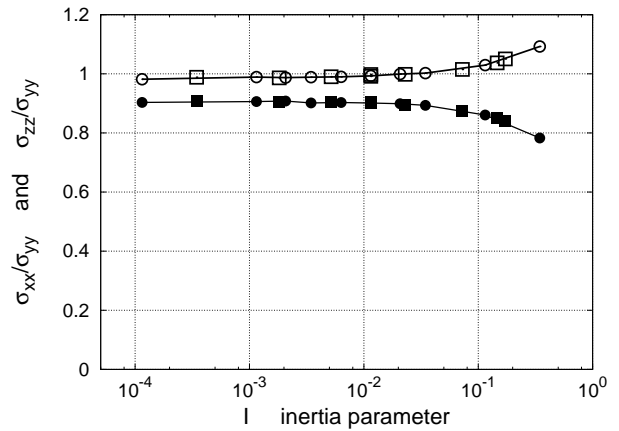}
\caption{The ratios of normal stresses are shown as the functions of the inertia
  parameter $I$. Open and closed symbols stand for
  $\sigma_{xx}/\sigma_{yy}$ and $\sigma_{zz}/\sigma_{yy}$,
  respectively. Circles and squares have the same meaning as in
  Fig.~\ref{Phi}.
  \label{normstress}}
\end{figure}

Thus we conclude that in case of our virtual material the inertia number
$I$ is indeed the key control parameter and it determines the solid
fraction uniquely in homogeneous plane shear. Exactly the same holds for
the effective friction $\mueff$ and other ratios of the elements of the
stress tensor which are also unique functions of $I$ as shown in
Fig.~\ref{mueff} and Fig.~\ref{normstress}.

In Figs.~\ref{Phi}, \ref{mueff} and \ref{normstress}, where the $I$ axis is
in logarithmic scale, all the plotted quantities reach a plateau for small
inertia numbers. This plateau region can be regarded approximately as the
quasi static region. Its upper bound, i.e. the largest quasi static value
of $I$, depends on the required accuracy. The ideal quasi static flow is,
in fact, a mathematical limit case of $I \rightarrow 0$.  The homogeneous
and stationary plane shear in the quasi static limit can be characterized
by the well defined critical values:
\begin{equation}
\mucrit = \lim_{I \rightarrow 0} \mueff(I)
\,\,\,\,\,\,\,\,\,\, \text{and}
\,\,\,\,\,\,\,\,\,\, 
\Phicrit = \lim_{I \rightarrow 0} \Phi(I) \, .
\end{equation}
In our simulations we find that the deviations of $\Phi$ and $\mueff$ from
the corresponding critical values are proportional to $I$ for small inertia
numbers (insets in Figs.~\ref{Phi}, \ref{mueff}).

An interesting feature of the normal stresses is shown in
Fig.~\ref{normstress}. Regarding the $xy$ shear plane the values of normal
stresses $\sigma_{xx}$ and $\sigma_{yy}$ are very close to each other
provided the system is sheared quasi statically. In perpendicular
direction, however, the normal stress $\sigma_{zz}$ turns out to be lower
about $10$ percent. This weak stress response in side direction might be an
important factor in real three dimensional flows and lead to a non-trivial
stress distribution throughout the system.

In this section we examined how the quasi static critical state (in short
critical state) can be approximated by low inertia numbers. The simulations
presented in the rest of the paper remain as near as possible to this ideal
quasi static limit. This effort is bounded by the available computational
capacity because lower inertia numbers require more calculations.

\subsection{The effect of contact friction and size distribution on the
  critical state}

\label{frictionsection}

In this section we study how the critical state differs for different
materials. The materials used in our simulations differ either in the
coefficient of contact friction $\mu$ or in the size distribution of the
spherical grains. We use either a polydisperse system similarly to the
previous section, i.e. with grain radii chosen uniformly between $1.0$ and
$1.3$, or a monodisperse system with grain radius $1.0$. The inertia
parameters are set to $5.2*10^{-3}$ and $4.5*10^{-3}$ for polydisperse and
monodisperse simulations, respectively. The systems consist of $5000$
grains each and are tested with various values of $\mu$. The properties of
the critical state are measured and averaged over a typical total shear
strain $\gamma \approx 30$.

\begin{figure}[tb]
\includegraphics[scale=0.65]{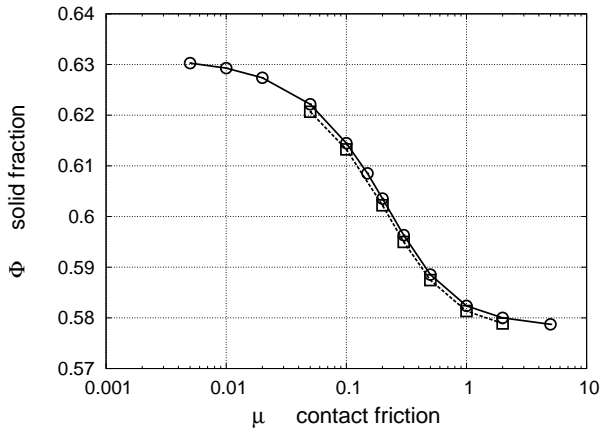}
\caption{The solid fraction $\Phi$ is plotted against the friction
  coefficient of the grain-grain contacts. The data shown by circles and
  squares are obtained in polydisperse and monodisperse systems,
  respectively. 
  \label{muPhi}}
\end{figure}
\begin{figure}[tb]
\includegraphics[scale=0.65]{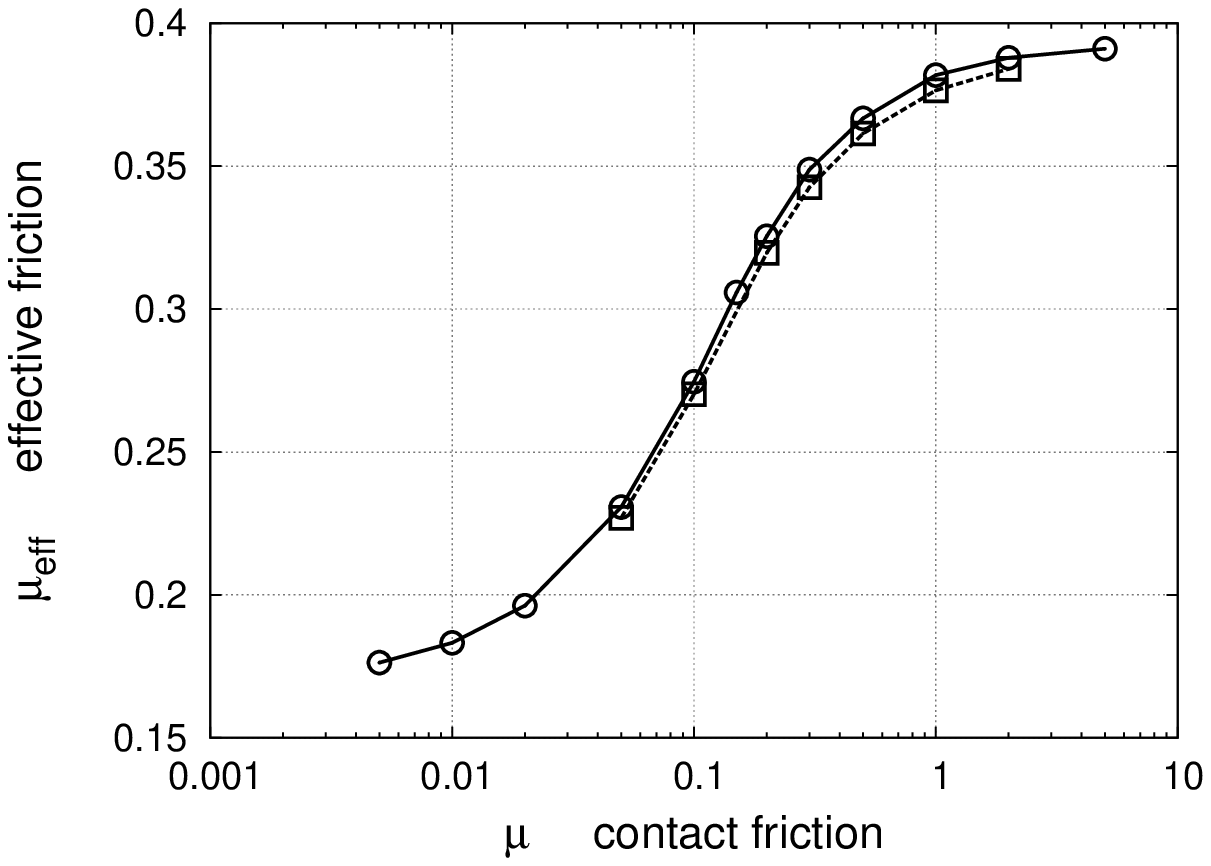}
\caption{The effective friction $\mueff$ is plotted against the friction
  coefficient of the grain-grain contacts. The data shown by circles and
  squares are obtained in polydisperse and monodisperse systems,
  respectively. 
  \label{mumueff}}
\end{figure}
\begin{figure}[tb]
\includegraphics[scale=0.65]{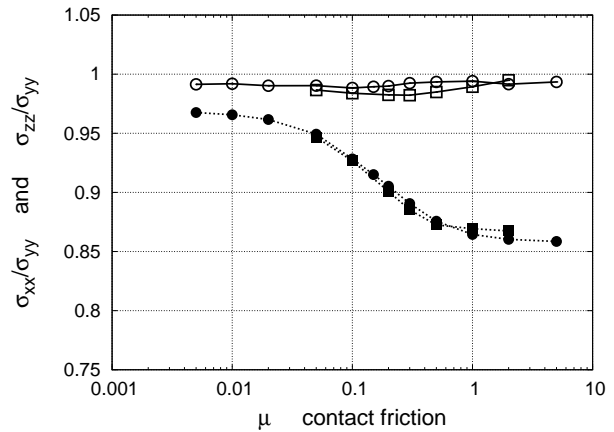}
\caption{The ratios of normal stresses are plotted against the friction
  coefficient of the grain-grain contacts. Open symbols show
  $\sigma_{xx}/\sigma_{yy}$ while closed symbols denote
  $\sigma_{zz}/\sigma_{yy}$. The data distinguished by circles and
  squares are obtained in polydisperse and monodisperse systems,
  respectively. 
  \label{munormstress}}
\end{figure}

In Fig.~\ref{muPhi}, \ref{mumueff} and \ref{munormstress} we show the
behavior of the quantities $\Phi$, $\mueff$ and the normal stress
ratios. Because the applied inertia parameters are relatively low, these
measured values can be regarded as approximate critical values. Actually,
the deviations between the data shown here and the exact critical values
can be estimated based on the results of the previous section. For example
the exact curve $\mucrit(\mu)$ is expected to be slightly lower than the
values of $\mueff$ shown in Fig.~\ref{mumueff} and the estimated deviation is
only around $6 * 10^{-3}$.

It it astonishing how little difference the polydispersity makes. It would
clearly be a different situation if we had much larger fluctuations in the grain
sizes. However, the level of polydispersity examined here has basically no
impact compared to a system of identical beads neither on the density nor
on the stress tensor.

The critical state, on the other hand, is very sensitive to the
inter-particle friction coefficient $\mu$. For example when changing $\mu$
the critical solid fraction (Fig.~\ref{muPhi}) explores almost the whole
range between the two limits $\Phi_{\text{RLP}}$ and $\Phi_{\text{RCP}}$
that are commonly attributed to the random loose and random close packings.

The fact that the effective friction depends strongly on $\mu$ is important
in the present scope. This property will be exploited later to study the
problem of the collective rheology.

We find (Fig.~\ref{munormstress}) that the normal stresses in the shear
plane ($\sigma_{xx}$ and $\sigma_{yy}$) remain approximately equal for the
whole range of $\mu$ as we saw this for one particular case in section
\ref{inertiasection}. Also the normal stress in side direction
$\sigma_{zz}$ appears to be weaker than the other two. However, the
friction $\mu$ has a significant influence on this weakening effect: While
for large $\mu$ the weakening is about $14$ percent, it drops down to $3$
percent with vanishing friction coefficient.

\section{Beyond the homogeneous case}

\subsection{The shear profile}

So far we focused on the homogeneous plane shear and examined the
properties of the critical state. In this section we turn to the case of
an inhomogeneous shear flow that is still relatively simple, stationary and
quasi static. We show that the behavior can not be interpreted based on the
critical state of the material.

When our shear cell is filled with one material as in the previous section,
the flow is indeed homogeneous. We find e.g. that the local stress tensor
averaged over time is the same everywhere in the shear cell. The time
averaged local velocity of the flow is parallel to the $x$ axis and the
speed is proportional to the $y$ coordinate. Thus the local shear rate is
independent of the position.

The inhomogeneity is introduced into the flow by using two different
materials. The upper half of the shear cell ($y>L_y/2$) is filled with
grains of contact friction $\muup$ that is chosen to be larger than the value
$\mulo$ that is used in the lower half ($y<L_y/2$). We recall that there is
no gravity in the simulation, ``up'' refers only to the orientation of the
$y$ axis. Apart from the coefficient of the contact friction there is no
difference between the two materials.

The main question is how the shear strain is distributed throughout the
system. The shear cell still remained translation invariant in $x$ and $z$
directions and no local property is expected to depend on these
coordinates. We will focus on the $y$ dependence of the measured quantities,
especially of the local shear strain. For such measurements the system is
divided into thin layers of thickness $\Delta y$ parallel to the $xz$ plane.
When the value of a measured quantity is reported later as the function
of $y$ it means that it is calculated for each layer, averaged over $x$,
$z$ and time over the whole range and over $y$ within the given layer.

To set up and study the above suggested system (or an equivalent one) in
real experiments might be very difficult. However, this shear cell suits
well to computer simulations and is very instructive regarding the problem of
collective rheology. As mentioned before, our goal is to avoid unnecessary
complications and study the simplest possible system where the problem can
be observed. The advantage of our system over traditional shear cells (like
Couette cell, cylindrical and straight split bottom cells) is manifold: (i)
the layers that slide next to each other in the shear flow are not curved
but are straight planes \cite{Depken06}, still stationary flow can be
maintained. (ii) Due to the high degree of symmetry local quantities are
not multivariate functions. They depend only on one parameter, the
coordinate $y$. (iii) Due to mechanical equilibrium the local shear stress
$\sigma_{xy}$ and the normal stress $\sigma_{yy}$ are bound to be constant
(do not depend on $y$). Thus the spatial stress distribution is a priori
very simple.

\begin{figure}[tb]
\includegraphics[scale=0.22]{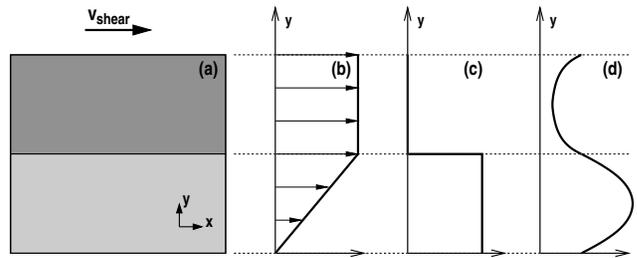}
\caption{(a) Sketch of the shear cell. The material in the upper part has
  larger coefficient of contact friction than that in the lower part. (b)
  and (c) The velocity $v(y)$ and the corresponding shear strain is
  shown, respectively, that are suggested by the relation $\mueff(I)$
  obtained for homogeneous plane shear. (d) For comparison, the shear strain is
  drawn qualitatively that is observed in the computer simulations.
  \label{schematicshearprofile}}
\end{figure}

How does such a system deform? What we could naively expect based on the
results of the homogeneous case is the following. Let us consider the lower
material first which is easier to shear. As $\sigma_{xy}/\sigma_{yy}$ does
not depend on $y$ the apparent effective friction is constant throughout
the lower material and so is the local inertia parameter
(Fig.~\ref{mueff}). This gives a constant shear rate $\dot\gamma(y)$ and a
linear velocity profile in this region
(Fig.~\ref{schematicshearprofile}). As the inertia parameter is kept very
small the stress ratio $\sigma_{xy}/\sigma_{yy}$ must be close to the
plateau value of the effective friction of the lower material
$\mu_{\text{crit,lo}}$. The same stress ratio must be valid also in the
upper part of the system. However, this stress ratio is not enough to cause
shear deformation at any rate in the upper material because here the
contact friction and therefore also the critical value
$\mu_{\text{crit,up}}$ is higher than in the lower material
(Fig.~\ref{mumueff}). Thus the overall stress ratio
$\sigma_{xy}/\sigma_{yy}$ is smaller than the minimum value
($\mu_{\text{crit,up}}$) that would be needed to maintain shear flow. This
naive argument leads us to the conclusion that the upper material does not
deform at all. The local shear rate $\dot\gamma(y)$ is constant in the
lower part and zero in the upper one.

This is in contrast to our numerical measurement where different behavior is
found (Fig.~\ref{schematicshearprofile}d). The observed flow does not
support the assumption that there is a one to one relation between the
local effective friction and the local inertia parameter. The function
$\mueff(I)$ that has been found in the homogeneous case is clearly not
valid here.

Regarding the inhomogeneous case, we present three computer simulations
$A$, $B$ and $C$ and the parameters of the simulations are as follows. The
systems are monodisperse (every grain radius is $1.0$), include $5000$
grains each, and are subjected to a typical global shear strain
$\gamma_{\text{glob}} = \Delta l_{\text{shear}} / (2 L_y) \approx 100$. The
driving shear rate $\dot\gamma_{\text{glob}}$ and the vertical pressure
$\sigma_{yy}$, which are kept constant in time, are chosen to be quasi
static: The corresponding global inertia parameter $I_{\text{glob}}$ is $5
* 10^{-4}$ in each case. In system $A$ and $B$ the pair of contact
frictions $(\mulo,\muup)$ is set to $(0.1,0.2)$ while for system $C$ the
contrast is larger: $(0.1,0.5)$. In case of $A$ and $C$ the two materials
occupy half and half of the shear cell as described before. This is
different for system $B$ where the $y$-position of the interface is set to
$0.7 \, L_y$ thus the width of the region of small (large) friction is $70$
($30$) percent of the width $L_y$ of the shear cell.

During the shear flow grains of the two materials could, in principle,
diffuse and mix. This effect is not in the present focus, however, in the
long run (in the time scale of mixing) would significantly alter the shear
profile. To avoid this interference we exploit the possibility provided by
the computer simulation and switch off the mixing effect as follows. If a
grain-grain contact is located below or above the interface then its
friction coefficient is set to $\mulo$ and $\muup$, respectively. Thus
contact friction is a property of the position and not of the grains in
these simulations which results in a sharp interface between the two
materials during the flow. In real experiments mixing can not be avoided in
this way. We note, however, that due to separation of time scales such
interfaces can exist long enough and mixing does not necessarily become an
issue \cite{Borzsonyi09,Knudsen09}.

\begin{figure}[tb]
\includegraphics[scale=0.65]{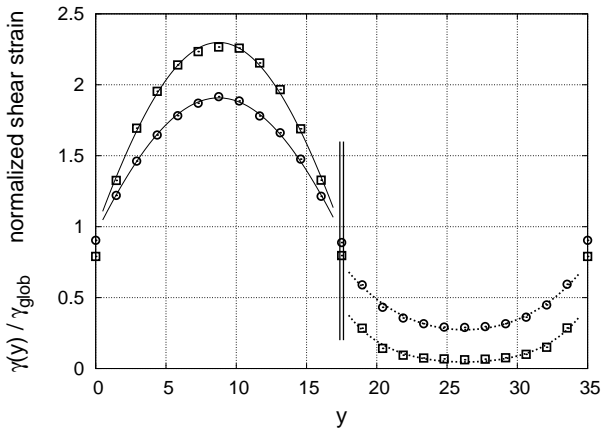}
\caption{The normalized shear strain is shown for systems $A$ (circles) and
  $C$ (squares) as the function of the vertical position $y$. Both systems
  have the same position of the interface indicated by the double vertical
  line. The pair of contact frictions ($\mulo$,$\muup$) used below and
  above the interface is ($0.1$,$0.2$) for $A$ and ($0.1$,$0.5$) for
  $C$. The data are fitted by cosine (hyperbolic cosine) functions shown by
  the solid (dotted) lines.
  \label{shearprofileAC}}
\end{figure}
\begin{figure}[tb]
\includegraphics[scale=0.65]{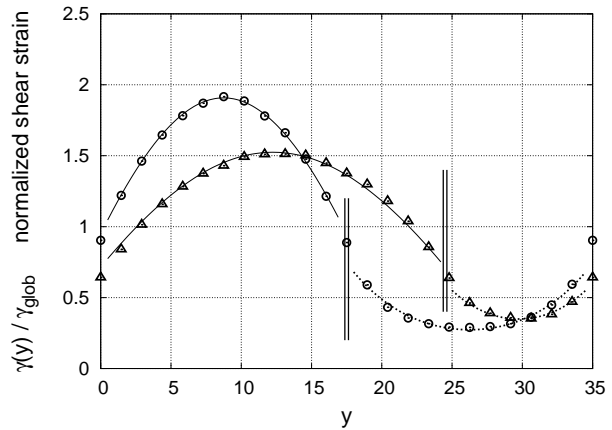}
\caption{The normalized shear strain is shown for systems $A$ (circles) and
  $B$ (squares) as the function of the vertical position $y$. Both systems
  have the same pair of contact frictions ($\mulo$,$\muup$) used below and
  above the interface: ($0.1$,$0.2$). They differ only in the position of
  the interface as indicated by the double vertical
  lines. The data are fitted by cosine (hyperbolic cosine) functions shown by
  the solid (dotted) lines.
  \label{shearprofileAB}}
\end{figure}

The data on the observed shear profiles are shown in
Figs.~\ref{shearprofileAC} and \ref{shearprofileAB} where the local shear
strain is plotted divided by the global shear strain that was applied to
the shear cell ($\gamma(y)/\gamma_{\text{glob}}$). These data are equivalent
to the normalized shear rate $\dot\gamma(y)/\dot\gamma_{\text{glob}}$ and
as long as the flow is quasi-static the observed curves are independent of
the driving shear rate $\dot\gamma_{\text{glob}}$. In each system the
steady state involves shear deformation of both the lower and the upper
materials.  These simulations demonstrate clearly that various values of the
local shear rate are possible for the same local stress. The measured
effective friction $\mueff = \sigma_{xy}/\sigma_{yy}$, which can be
interpreted as the shear resistance that the inhomogeneous system collectively
develops against the external driving, turns out to be very similar in all
cases $A$, $B$ and $C$: $\mueff = 0.275 \pm 0.003$. This value is somewhat
larger then the estimated critical friction for the lower material:
$\mucrit(\mu=0.1) \approx 0.264$ and smaller than those for the two
high-friction materials: $\mucrit(\mu=0.2) \approx 0.314$ and
$\mucrit(\mu=0.5) \approx 0.356$.

\subsection{Tentative description}

How can we interpret the behavior presented in Figs.~\ref{shearprofileAC}
and \ref{shearprofileAB}? Why does, in the first place, the region of
$\muup$ flow at all for shear stresses that are too weak to maintain flow
in the homogeneous case? Presumably, the material of $\muup$ remains solid
for such stresses only if there is no flow around. However the same
material can not retain the same static shear stress in the inhomogeneous
situation because the lower part of the system flows already. This flow
provides a kind of noise in the force network of the upper material. Thus
the region that is expected to be solid is exposed to permanent agitation
which must be quite strong near the interface. Under these noisy conditions
the inner fabric might fail from time to time which results in a creep
motion in the direction of the local shear stress no matter how weak this
shear stress is. The key concept must be in our view that local deformation
represents also the source of agitation \cite{Nichol10}. Even without
knowing exactly the nature of the agitation effect it seems to be plausible
that this mechanism is responsible for the observed quasi static rheology
in various systems. The agitations arising from the nearby flow was
expected to rule out static shear stresses also in the experiment
\cite{Nichol10} where sand was turned into a zero yield stress fluid.

Let us set up a tentative model in the spirit of the above picture to
describe the rheology seen in the previous section. We consider an array of
narrow parallel layers perpendicular to the $y$ axis with their width $\xi$
much smaller than the system size.  The shear strain of the $i$th layer is
denoted by $\gamma_i$. The value $\sigma_{yy}$, which is a fixed constant
in the simulation, is taken as the unit of stress and we express the shear
stress in this unit $\tau = \sigma_{xy}/\sigma_{yy}$. We assume about the
local shear strain $\gamma_i$ that it depends on two factors, on the local
shear stress and on the amount of agitation that the $i$th layer
receives. Furthermore, the deformation of the neighboring layers serves as
the source of agitation, i.e. the amount of agitation received by the layer
$i$ is the function of $\gamma_{i-1}$ and $\gamma_{i+1}$ only. We put this
in the following form:
\begin{equation}
\gamma_i = (\gamma_{i-1} + \gamma_{i+1}) f(\tau) \, ,
\label{eqmodelstart}
\end{equation} 
where the plausible assumption is taken that the caused deformation
$\gamma_i$ is proportional to the deformation $\gamma_{i-1} + \gamma_{i+1}$
that creates it. About the effect of $\tau$ we assume only that it is given
by a monotonous increasing function $f(\tau)$ that may be different for
different materials. It means that for the same amount of agitation
the resulting strain grows with the shear stress. Eq.~\ref{eqmodelstart} is
equivalent to the discrete equation:
\begin{equation}
\frac{\gamma_{i-1} + \gamma_{i+1} - 2 \gamma_i}{\xi^2} 
= 
\frac{2}{\xi^2} \left[\frac{1}{2 f(\tau)}-1 \right] \gamma_i \, .
\label{eqdiscrete}
\end{equation}
In a continuum description (on length scales larger than $\xi$) this
corresponds to the following differential equation for $\gamma(y)$:
\begin{equation}
\frac{\partial^2 \gamma}{\partial y^2} = C(\tau) \gamma \, ,
\label{eqdiff}
\end{equation}
where $C(\tau)$ is a material dependent function of the shear stress
\begin{equation}
C(\tau)=\frac{2}{\xi^2} \left[\frac{1}{2 f(\tau)}-1 \right] \, .
\end{equation}

The aim of Eq.~\ref{eqdiff} is to apprehend the shear profile in the
studied flow which is quasi static and stationary. The structural form of
the equation seems to be appropriate: it does not involve time and time
derivatives which is a nice feature for a quasi static model, furthermore,
if $\gamma(y)$ is a solution then $\lambda \gamma(y)$ is a solution as well
for any number $\lambda$ as it should be in steady state.

Let us consider the homogeneous situation first where $\gamma(y)$ is
independent of $y$ and $\tau$ is equal to the critical stress ratio
$\taucrit$. Then the left hand side of Eq.~\ref{eqdiff} is zero and the
value of $C(\tau)$ must vanish for $\taucrit$. It also gives
$f(\taucrit)=1/2$. This has an important consequence beyond the homogeneous
case, namely, that the sign of $C$ is different below and above the
critical shear stress. Because $f(\tau)$ is monotonous increasing:
\begin{equation}
\begin{split}
C(\tau) > 0 \,\,\,\,\,\,\,\,\,\,\,\, \text{if} \,\,\,\,\,\,\,\, \tau <
\taucrit \, ,
 \\
C(\tau) < 0 \,\,\,\,\,\,\,\,\,\,\,\, \text{if} \,\,\,\,\,\,\,\, \tau >
\taucrit \, .
\end{split}
\end{equation}

Eq.~\ref{eqdiff} allows quasi static flow also for shear stresses
below and above the critical value $\taucrit$ and it makes also possible
that different parts of the material exhibit different extent of
deformation under the same local stress. This is in contrast to the naive
picture we deduced alone from the homogeneous simulations of section
\ref{inertiasection}.

The shear resistance that the material sets up against the external driving
may be smaller or also larger than the critical value seen in the
homogeneous plane shear. Thus not only softening but also hardening is
possible compared to this reference state.
Eq.~\ref{eqdiff} tells us that the deviation from the critical stress results in
a curved shear profile $\gamma(y)$. The curvature is positive (negative)
for smaller (larger) values of $\tau$.

$\tau$ is constant in space in our shear cell for both the homogeneous and
the inhomogeneous simulations. Therefore $C(\tau)$ is always a constant
throughout one material and for such a region the functional form of
$\gamma(y)$ can be easily determined from Eq.~\ref{eqdiff}. 
\begin{equation}
\gamma_{\text{lo}}(y) = \Alo \, \text{cos}\left[ k_{\text{lo}}\left(y-y_c\right)\right]  
\end{equation}
for the lower materials in Figs.~\ref{shearprofileAC} and
\ref{shearprofileAB} where $y_c$ is the actual middle position of the given
region (where we utilized the mirror symmetry of the setup with respect to
the symmetry plane of position $y_c$). The parameter $k_{\text{lo}}$ is
given by $C_{\text{lo}}(\tau) = - k_{\text{lo}}^2$, where $C$ is indeed
negative as $\tau>\taucrit$ and the curvature of $\gamma$ is also
negative. For the upper materials where in each case the overall shear
stress $\tau$ is smaller than $\taucrit$ of the given material the value
$C_{\text{up}}(\tau)$ is positive and the corresponding solution has
positive curvature:
\begin{equation}
\gamma_{\text{up}}(y) = \Aup \, \text{cosh}\left[ k_{\text{up}} \left(y-y_c\right)\right] \, ,  
\end{equation}
where $k_{\text{up}}^2 = C_{\text{up}}(\tau)$. 

The above model gives nice interpretation of some features of the quasi
static behavior. As mentioned already both softening and hardening is
possible. Hardening is interpreted that the $i$th layer gets less amount of
agitation per unit shear strain than it needs for homogeneous
deformation. The reduced level of agitation is connected to negative
curvature of the shear profile $\gamma(y)$ and makes the material harder to
shear ($\tau > \taucrit$). Softening compared to the reference critical
state is the opposite case which involves enlarged level of agitations,
positive curvature of $\gamma$ and smaller shear stress than the critical
one. The model also does a good job in predicting the functional form of
the shear profile as the functions cosine and hyperbolic cosine match the
observed data of $\gamma(y)$ very well (Figs.~\ref{shearprofileAC} and
\ref{shearprofileAB}).

There are of course many questions left open in our quasi one dimensional
shear cell which are important also to understand quasi static flows in
general. For example we can not calculate the parameters $k_{\text{lo}}$ and
$k_{\text{up}}$ as we do not know the function $C(\tau)$. The prediction of
ratio $\Alo/\Aup$ is also missing here. Therefore all these quantities are
fit parameters in Figs.~\ref{shearprofileAC} and \ref{shearprofileAB}. The
predicted curves are fitted only inside the given material, data points at
the boundaries are not included because they represent an average shear
strain in the vicinity of the interface involving both materials.

In the above picture the shear stress $\tau$ determines $k$ but not the
amplitude. In the lower material $\Alo$ depends only on the amount of
agitation that this region gets through the boundaries, that is, on the
agitation that is generated by the other material and transmitted through
the two interfaces. In turn, $\Aup$ is determined by the transmitted
agitation through the interfaces in the other direction towards the upper
material. There is no reason why $\gamma(y)$ should be continuous at such
material interfaces. At this point we are not able to tell the boundary
condition for $\gamma(y)$ since not enough is known about the agitation
mechanism.

\subsection{The solid fraction}

Let us consider the inhomogeneous shear flow in our shear cell and the
local state of the material at a given $y$ position. How does this state
differ from the critical state of the same material obtained in the
homogeneous case? We argued that the two situations are somehow different
in two quantities, in the amount of the hypothetical agitation per unit shear
strain and in the more specific stress ratio $\sigma_{xy}/\sigma_{yy}$
under which the flow occurs. Here we would like to emphasize and
demonstrate it more clearly that the two states are different, even in
the inner structure of the material. 

\begin{figure}[tb]
\includegraphics[scale=0.65]{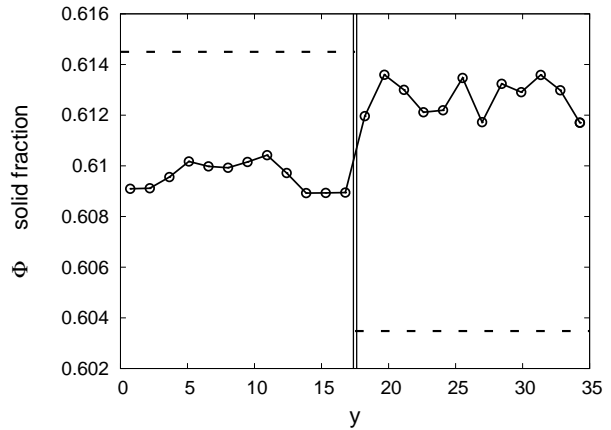}
\caption{The local solid fraction is plotted for simulation $A$ as the
  function of the vertical position $y$. The double vertical line indicates
  the position of the interface between the lower and the upper
  materials. For both materials the horizontal dashed lines show the values
  $\Phicrit$, i.e. the corresponding solid fractions in case of a quasi static
  homogeneous plane shear.
  \label{yPhi}}
\end{figure}

This can be well seen if we plot the local solid fraction $\Phi(y)$
measured in the inhomogeneous flow. Fig.~\ref{yPhi} shows the data for
simulation $A$, where also the critical solid fractions of the two
materials are shown for comparison that were obtained by homogeneous plane
shear. The lower material exhibit a solid fraction in this stationary and
quasi static flow that is smaller than its well defined critical value. The
upper material does the opposite, it gets denser compared to its critical
density. We interpret the phenomenon as follows. The shear deformation
tends to create pores between the grains while agitations, a sort of random
shaking, tends to destroy these pores and drive the material towards a
relatively dense structure. Compared to the homogeneous situation the upper
material is deformed in an environment with increased level of agitation
therefore the densification works better. For the lower material the
reduced level of agitation results in looser structure than in the critical
state.

Whether the speculation about the agitation level is right or not, the
effect is clearly shown that the lower material is packed looser than in
its critical state while the upper material gets denser than its own
critical density. This effect is so strong that actually the order of the
densities of the two materials is reversed. This is surprising because it
is expected that spherical grains with larger contact friction have
lower solid fraction. Here we find the opposite within one system in a
stationary flow.

\section{Summary}

This paper is subjected to the quasi static flow of granular materials. We
argued that quasi static rheology is a collective phenomenon that can not
be interpreted locally as a simple plane shear. In general, the collective
flow can exhibit such local states of the material that do not exist at all
in homogeneous plane shear tests. Therefore it is difficult to isolate
these states and study them separately. Even if we restrict the study to
quasi static and stationary shear flows the physical properties of a
mesoscopic piece of bulk material can be very different from the properties
of the well defined critical state (e.g. density, shear resistance). The
collective rheology can not be understood merely based on a constitutive
relation that connects the local stress to the local strain and strain
rate.

We analyzed a rather simple shear cell that was designed to
demonstrate these fundamental problems of collective granular flows. We
used three dimensional computer simulations and analyzed the spatial
distribution of the stress and strain. We discussed a physical picture that
may help to understand the observed flow in which the local deformation
agitates the surrounding material and therefore creates deformation. We set
up a tentative continuum model that essentially states that the local shear
strain $\gamma$ is proportional to its Laplacian $\Delta \gamma$ and the
factor between them is determined by the local stress. The shear profile
predicted by the model is in nice agreement to the observed deformation in our
quasi one dimensional system.

\begin{acknowledgments}
The author acknowledges support by the Bolyai J\'anos research program and
by the Hungarian Scientific Research Fund (grant No. PD073172).
\end{acknowledgments}

\bibliography{Ref}

\end{document}